\documentclass[floatfix,nofootinbib,10pt,notitlepage]{revtex4-1}

\usepackage{graphicx}
\usepackage{amsmath,amssymb,amsfonts}
\usepackage{bbm}

\newcommand{\Br}{{\bf r}}
\newcommand{\Bu}{{\bf u}}
\newcommand{\Bv}{{\bf v}}
\newcommand{\Btau}{\boldsymbol{\tau}}
\newcommand{\BF}{{\bf F}}
\newcommand{\BB}{{\bf B}}
\newcommand{\BD}{{\bf D}}
\newcommand{\Bn}{{\bf n}}
\newcommand{\Bw}{{\bf w}}

\newcommand{\BU}{{\bf U}}
\newcommand{\BT}{{\bf T}}
\newcommand{\Bsigma}{\boldsymbol {\sigma}}

\newcommand{\Bdelta}{\boldsymbol{\delta}}

\newcommand{\Bs}{B_{\rm s}}
\newcommand{\BG}{{\bf G}}

\newcommand{\ie}{{i.e., }}
\newcommand{\kT}{k_{\rm B}T}

\newcommand{\pd}{\partial}

\newcommand{\SusD}{{\rm 3D}}
\newcommand{\MemD}{{\rm 2D}}
\newcommand{\BC}{{\bf C}}

\begin{document}

\title{Many-particle mobility and diffusion tensors for objects in viscous sheets}

\author{Yulia Sokolov}
\affiliation{Raymond \& Beverly Sackler School
  of Chemistry, Tel Aviv University, Tel Aviv 6997801, Israel}

\author{Haim Diamant}
\affiliation{Raymond \& Beverly Sackler School of Chemistry, Tel Aviv University, Tel Aviv 6997801, Israel}
\email{hdiamant@tau.ac.il}
\date{\today}

\begin{abstract}
  
We derive a mobility tensor for many cylindrical objects embedded in a
viscous sheet.  This tensor guarantees a positive dissipation rate for
any configuration of particles and forces, analogously to the
Rotne-Prager-Yamakawa tensor for spherical particles in a
three-dimensional viscous fluid. We test our result for a ring of
radially driven particles, demonstrating the positive-definite
property at all particle densities. The derived tensor can be utilized
in Brownian Dynamics simulations with hydrodynamic interactions for
such systems as proteins in biomembranes and inclusions in
free-standing liquid films.

\end{abstract}

\maketitle

\section{Introduction}
\label{sec_intro}

Many systems in nature are based on thin sheets of a viscous fluid.
The main example is biomembranes \cite{SD}. Other examples are soap
films \cite{Weeks}, liquid crystalline films \cite{Nguyen,Qi} and
monolayers at fluid-fluid interfaces \cite{Prasad}. Many studies have
been devoted to the in-plain dynamics of such systems, some of which
are reviewed in Ref.~\citenum{Brown}. These studies include, in
particular, the derivation of the self-mobility of an isolated
cylindrical particle in a viscous sheet \cite{Saffman,Hughes}, the
pair-mobility of two such objects \cite{Naomi}, as well as the
hydrodynamic kernel associated with the flow due to a point-force
\cite{LevineDyn,LevineMob}. We briefly review these results below. In
addition, various numerical schemes were developed to deal with the
complex membranal dynamics \cite{Brown,Naji,Camley,Noruzifar}.

The present work relates to the dynamics of multiple mobile objects
within viscous sheets. As shown below, there are stability issues with
the currently used many-particle mobility tensor, arising from the
fact that it is not positive-definite. A similar problem is well-known
in the case of particles in three-dimensional (3D) suspensions
\cite{Zwanzig}, and was famously solved by Rotne and Prager \cite{RP}
and Yamakawa \cite{Yamakawa}. Here we solve it for the analogous
quasi-two-dimensional (2D) case of viscous sheets.

We assume the usual limit of overdamped dynamics (vanishing Reynolds
number). In this limit the response of the objects to forces is
linear, instantaneous, and can be characterized by mobility
coefficients. In biological systems these conditions generally apply
\cite{SD,Saffman}. We focus on the translation of the objects and do
not treat rotation. Along the text, we refer to mobilities $\BB$
rather than diffusivities $\BD$. At equilibrium, the two are related
by the thermal energy, $ֿ\BD=\kT \BB$ (the Einstein relation).

We proceed with a brief summary of known results for the mobility of a
single particle and a pair of particles, in 3D and 2D. Then we refer
to many-particle dynamics and describe the problem of negative
mobility and its correction in 3D. In Sec.~\ref{sec_PDT} we derive a
2D mobility tensor, which is positive-definite by construction. We
examine the derived tensor on two cases, testing its positiveness. In
Sec.~\ref{sec_disc} we discuss the results and possible extensions and
applications. Appendices A, B, and C provide additional information
which may be useful for future simulations.

\subsection{An isolated particle}

The self-mobility of a particle, in general, characterizes its linear
velocity response to the force applied to it,
\begin{equation}
\Bv_{\alpha}=\BB_{{\rm s},\alpha\beta}\BF_{\beta},
\label{Bs_general}
\end{equation}
where Greek indices denote spatial coordinates, and we sum over
repeated indices. In a 3D fluid of viscosity $\eta$, the self-mobility of a sphere of radius $a$ is given by Stokes' formula \cite{StokesHappelBrenner},
\begin{equation}
  ^{\SusD}\BB_{{\rm s},\alpha\beta}={}^{\SusD}\Bs\Bdelta_{\alpha\beta}, \ \ \
 ^{\SusD}\Bs=(6\pi\eta a)^{-1}.
\label{Stokes}
\end{equation}

\begin{figure}
\centering
 \includegraphics[width=0.4\columnwidth]{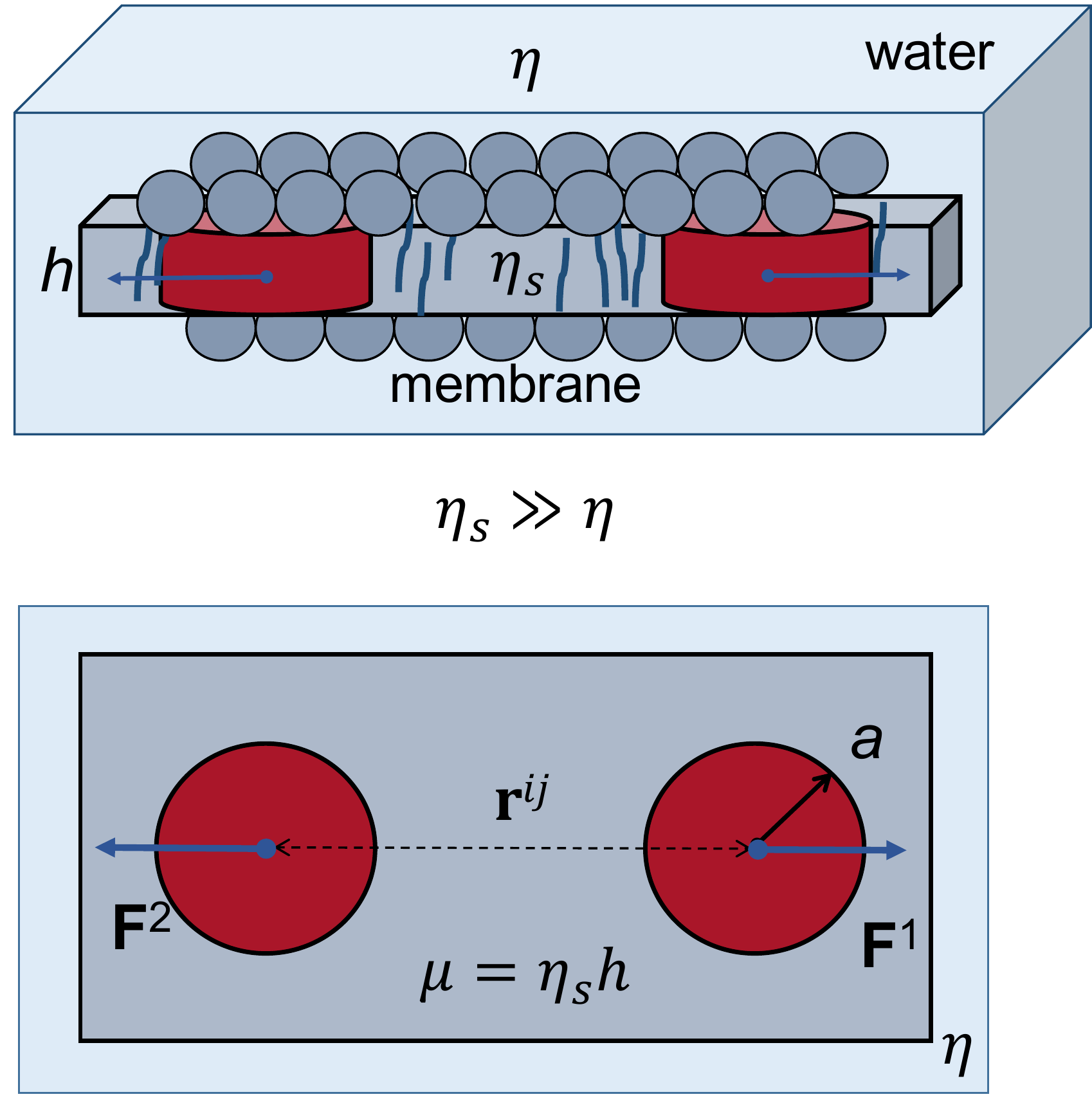}
\caption{Illustration of a membrane of viscosity $\eta_s$ and
  thickness $h$, consisting of amphiphilic lipids (lighter, grey
  beads), with two driven proteins modeled as cylinders of radius $a$
  (darker, maroon cylinders), immersed in a 3D surrounding fluid of
  viscosity $\eta$. }
\label{fig:membrane}
\end{figure}

In a viscous sheet of viscosity $\eta_{\rm s}$ and thickness $h$, the
self-mobility of a particle of sufficiently small size $a$ is in
general
\begin{equation}
  ^{\MemD}\BB_{{\rm s},\alpha\beta}={}^{\MemD}\Bs\Bdelta_{\alpha\beta},\ \ \
  ^{\MemD}\Bs = \frac{1}{4 \pi \mu} \left(\ln \frac{2}{\kappa a} -\gamma \right),
  \ \ \ \kappa a\ll 1.
\label{Bs}
\end{equation}
Here $\mu=\eta_{\rm s} h$ is an effective 2D viscosity, $\gamma\simeq 0.58$
is Euler's constant, and $\kappa^{-1} \gg a$ is an upper cut-off
length required to regularize 2D hydrodynamics. In the specific
example of the Saffman-Delbr\"uck model for a protein in a biomembrane
\cite{SD}, illustrated in Fig.~\ref{fig:membrane}, the self-mobility
is $^{\MemD}\Bs = (4\pi\mu)^{-1}[\ln(2\lambda/a)-\gamma]$. Here, the
cut-off length $\kappa^{-1}$ is the Saffman-Delbr\"uck length
$\lambda=\mu/(2 \eta)$, arising from the difference between the
viscosities of the membrane and the surrounding fluid. Similar
calculations for other viscous sheets and particle shapes all lead, in
the limit $\kappa^{-1} \gg a$, to a self-mobility similar to
Eq.~(\ref{Bs}), with varying definitions of the cut-off
$\kappa$. \cite{Evans,Stone,NajiMob,Daniels,Henle,Ramachandran}

\subsection{A pair of particles}

When two or more particles move within a viscous fluid, they do not move independently.
Mutual drag forces\,---\,hydrodynamic interactions\,---\,correlate their motions. For a single pair of particles Eq.~(\ref{Bs_general}) is generalized to
\begin{equation}
\Bv^i_{\alpha}=\BB^{ij}_{\alpha\beta} \left(\Br\right)\BF^j_{\beta},
\label{B_pair}
\end{equation}
where the Latin indices $i,j=1,2$ mark the particles, and
$\Br=\Br^2-\Br^1$ is the vector connecting their positions. Equation (\ref{B_pair}) 
describes the velocity response of each particle to the
forces acting both on it and on its partner. The diagonal blocks ($i=j$) of
the pair-mobility tensor give the self-mobility of each particle in the presence
of its partner, given their separation $\Br$, while the off-diagonal blocks ($i\neq j$) give
their coupling due to the hydrodynamic interactions.

Within the Stokeslet approximation, valid in the limit of large
separations compared to particle size ($r\gg a$), the particles are
considered arbitrarily small, resulting in
\begin{equation}
  \BB^{11}_{\alpha \beta} = \BB^{22}_{\alpha \beta} \simeq
  \BB_{{\rm s},\alpha\beta}, \ \ \ 
  \BB^{12}_{\alpha \beta} = \BB^{21}_{\alpha \beta} \simeq
  \BG_{\alpha \beta} (\Br),
\label{Bpair}
\end{equation}
where $\BG_{\alpha\beta}(\Br)$ is the velocity response of the fluid
at position $\Br^2$ to a point-force at $\Br^1$ (the Green's function of the flow
equations). Thus, in this limit, the diagonal block is just a self-mobility of an
isolated particle, and the off-diagonal block is the coupling mobility of two point-like
particles.

For a 3D suspension, the fluid's response is given by the Oseen tensor
\cite{Oseen},
\begin{equation}
  ^{\SusD} \BG_{\alpha\beta}(\Br) =\frac{1}{8 \pi \eta
    r} \left(\Bdelta_{\alpha \beta} + \frac{\Br_{\alpha}
    \Br_{\beta}}{r^2} \right),
\label{Oseen}
\end{equation}
and the interaction strength decays with distance as $1/r$. This relatively slow decay
corresponds to long-range hydrodynamic interactions between particles in suspensions.

For cylinders in a sheet, the analog of the Oseen tensor in the limit of  $r \ll \kappa^{-1}$ is \cite{Lubensky, Bussel,LevineDyn,LevineMob,Naomi}
\begin{equation}
   ^{\MemD}\BG_{\alpha\beta}(\Br) = \frac{1}{4 \pi \mu}
  \left[ \left(\ln \frac{2}{\kappa r} -\gamma -\frac{1}{2} \right)
    \Bdelta_{\alpha \beta} + \frac{\Br_{\alpha} \Br_{\beta}}{r^2}
    \right].
\label{Bpsd}
\end{equation}
Equations~(\ref{Bs}),
(\ref{Bpair}), and (\ref{Bpsd}) yield the pair-mobility
tensor of two cylindrical particles within the Stokeslet approximation.
The logarithmic decay with distance implies
much longer-range correlations than in a 3D fluid, 
up to distances of order $\kappa^{-1}$. As a result, local perturbations give rise
to strong nonlocal effects.

\subsection{Many-particle dynamics}

Let us consider an ensemble of many objects within a viscous sheet. 
In general, the velocity response of $N$ particles to the forces $\mathbb{F}=(\BF^1,\ldots\BF^N)$ acting on them, is given by a many-particle mobility tensor, according to
\begin{equation}
  \Bv^i_{\alpha}=\BB^{ij}_{\alpha\beta} \left(\mathbbm{r}
  \right)\BF^j_{\beta},
\label{B}
\end{equation}
where $i,j=1\ldots N$ are particle labels, and $\mathbbm{r}=
(\Br^1,..., \Br^N)$ is the configuration defined by the positions of
all particles. A simple way to construct an approximate
many-particle mobility tensor is to assume superposition of pair-mobilities.
The pair-mobility in the Stokeslet approximation ($|\Br^{ij}|\gg a$ for all pairs),
Eq.~(\ref{Bpair}), gives
\begin{equation}
  \BB^{i=j}_{\alpha \beta} = \BB_{{\rm s},\alpha\beta}, \ \ \ 
  \BB^{i \neq j}_{\alpha \beta} = \BG_{\alpha \beta} (\Br^{ij}),
\label{Bk}
\end{equation}
where $\Br^{ij}=\Br^j-\Br^i$.
Substitution of the expressions (\ref{Stokes}) and (\ref{Oseen}) for a 3D
suspension in Eq.~(\ref{Bk}) yields the Kirkwood-Riseman (KR)
tensor \cite{Kirkwood}. The 2D-analog of this tensor is
obtained by using the expressions (\ref{Bs}) and (\ref{Bpsd}) in Eq.~(\ref{Bk}) \cite{LevineMob}.

\subsection{Positive-definite mobility tensor}
\label{sec_positive}

If the mobility tensor is not positive-definite, there exist 
configurations, for which the resulting power imparted to the particles,
\begin{equation}
  \epsilon = \BF^i_\alpha\Bv^i_\alpha = \BF^i_\alpha
  \BB^{ij}_{\alpha\beta}(\mathbbm{r}) \BF^j_\beta,
\label{dissipation}
\end{equation}
is negative. However, in the viscous regime this power equals 
the energy dissipation rate in the fluid, which cannot be negative due to the second law of thermodynamics. The Stokeslet approximation fails to fulfill this requirement, as it becomes invalid once there are configurations with insufficiently large separations \cite{Zwanzig}. The problem emerges, in particular, in Brownian Dynamics simulations,
where configurations involving close-by particles are
inevitable. The KR tensor gives rise
to instability and non-physical dynamics in such simulations \cite{Zwanzig}.

Rotne and Prager \cite{RP}, and
Yamakawa \cite{Yamakawa}, calculated an improved tensor, which overcomes the problem
of the KR tensor in 3D. The derivation by Rotne and Prager
is based on an Ansatz for the stress tensor, taking into account the
finite size of the particles, and integrating the flow response over
their surfaces. This variational treatment yields
a diffusion tensor, which goes beyond the Stokeslet limit. Although the tensor is not
expected to be accurate for small separations, it (a) ensures a positive
dissipation rate for all particle and force configurations, and (b) converges to
the KR tensor for large separations. This Rotne-Prager-Yamakawa
(RPY) tensor is given by
\begin{eqnarray}
  ^{\SusD}\BB^{i=j}_{\alpha \beta} = ^{\SusD}\BB_{{\rm s},\alpha\beta}, \ \ \ 
  && ^{\SusD}\BB^{i \neq j}_{\alpha \beta} = ^{\SusD}\BG_{\alpha\beta}(\Br^{ij})
  + \BC(\Br^{ij}), \nonumber\\
  &&\BC(\Br)=\frac{1}{12 \pi \eta r} \frac{a^2}{r^2} \left(  \Bdelta_{\alpha \beta} - 3 \frac{\Br_{\alpha}\Br_{\beta}}{r^2} \right),
\label{BRPY}
\end{eqnarray}
where $^{\SusD}\BB_{\rm s}$ remains the Stokes mobility of
Eq.~(\ref{Stokes}), and $^{\SusD} \BG_{\alpha\beta}$ is the Oseen
tensor, Eq.~(\ref{Oseen}).  Compared to the KR tensor of
Eq.~(\ref{Bk}), the RPY tensor introduces a correction to second order
in $a/r$. In fact, Eq.~(\ref{BRPY}) is equivalent to a superposition
of pair-mobilities, corrected to order $(a/r)^2$, which was not noted
in the original paper.

Similarly, the negative mobility problem arises when using
 the 2D-analog of the KR tensor, obtained from Eqs.~(\ref{Bs}), (\ref{Bpsd}), and
(\ref{Bk}). Here as well, the Stokeslet approximation fails for short-distance configurations, 
as will be demonstrated below. In the following section we apply the Rotne-Prager construction to
the 2D case.

\section{2D Positive-definite mobility tensor (PDT)}
\label{sec_PDT}

We follow the line of argument of Ref.~\citenum{RP} while adapting it to
cylinders in a sheet. Since we restrict the discussion to distances
$r\ll\kappa^{-1}$, the analysis is essentially two-dimensional, \ie
the dynamics is confined to the plane of the sheet. The effect of the 3D surroundings 
is captured by the cut-off $\kappa$ only. Thus, from now on we deal with a two-dimensional problem and omit the superscripts 2D.

Consider a configuration $\mathbbm{r}$ of
cylindrical particles and forces
$\mathbb{F}$ acting on them. The resulting exact flow field
$\Bu(\Br)$, satisfying the boundary conditions on the perimeters of all
particles, is unknown. The stress tensor associated with it is 
$\Bsigma_{\alpha\beta}(\Br)=-p(\Br)\Bdelta_{\alpha\beta}+\Btau_{\alpha\beta}(\Br)$,
where $p(\Br)$ and $\Btau(\Br)$ are the exact pressure and
viscous-stress fields, with
\begin{equation}
  \Btau_{\alpha \beta} (\Br) =
  \mu \left[\partial_{\alpha} \Bu_{\beta}(\Br) +
  \partial_{\beta} \Bu_{\alpha}(\Br) \right].
\label{tau}
\end{equation}
Even though the exact stress tensor is unknown, it can be approximated, provided
that the following basic properties are maintained. 
(a) Both $\Bsigma$ and $\Btau$ are symmetric tensors, $\Bsigma_{\alpha\beta}=\Bsigma_{\beta\alpha}$ and $\Btau_{\alpha\beta}=\Btau_{\beta\alpha}$. (b)
$\Btau$ is traceless, $\Btau_{\alpha\alpha}=0$. (c) The total stress
tensor is divergenceless (local forces acting on a fluid element balance to
zero since inertia is neglected),
\begin{equation}
  \pd_\beta\Bsigma_{\alpha\beta}(\Br) = -\pd_\alpha p(\Br) +
  \mu\pd_{\beta\beta}\Bu_\alpha(\Br) = 0.
\end{equation}
(d) For the same reason, the total force exerted by the 2D flow
on the perimeter of each particle exactly balances the external force
applied to it,
\begin{equation}
  a \int_0^{2\pi} d\theta
  \,\Bsigma_{\alpha\beta}[\Br^i+a\Bn^i(\theta)]\,\Bn^i_\beta(\theta) = \BF^i_\alpha,
\end{equation}
where $\Bn^i(\theta)$ is a unit normal to the perimeter of cylinder
$i$. Once the viscous stress is formulated, one can obtain
the mobility tensor by demanding that the total dissipation rate produced by the viscous flow must be equal to the power imparted to the particles by the external forces,
\begin{equation}
  \epsilon = (2\mu)^{-1} \int d^2r \,\Btau_{\alpha\beta}\Btau_{\beta\alpha} =
  \BF^i_\alpha \BB^{ij}_{\alpha\beta}(\mathbbm{r}) \BF^j_\beta.
\end{equation}
Note that the integration is over the fluid domain of the sheet, excluding the
areas of the particles.

The exact viscous stress tensor $\Btau$, evidently, is
positive-definite, producing a strictly positive dissipation rate
$\epsilon$ for any configuration. In linear
hydrodynamics it is also a {\em minimizer} of
$\epsilon$. \cite{Doibook} Any choice of a valid stress tensor satisfying the
requirements (a)--(d), is bound to produce a
dissipation rate above the minimum, $\bar\epsilon\geq\epsilon>0$, 
and thus be positive-definite and
correspond to a PDT.

Following these guidelines, we construct a stress tensor based on
the flow induced by a single forced particle, and define
\begin{eqnarray}
  \bar\Bsigma_{\alpha\beta}(\Br) &\equiv& -\bar{p}(\Br)\Bdelta_{\alpha\beta} + \bar\Btau_{\alpha\beta}(\Br)
  = \sum_i \Bsigma^i_{\alpha\beta}(\Br-\Br^i), \nonumber\\
  \bar{p}(\Br) &\equiv& \sum_i p^i(\Br-\Br^i),\ \ \ \bar\Btau_{\alpha \beta} \equiv \sum_i\Btau^i_{\alpha\beta} (\Br-\Br^i),
\label{stress}
\end{eqnarray}
where $\Bsigma^i$, $p^i$, and $\Btau^i$ are the total stress,
pressure, and viscous stress produced in the sheet by a single
particle located at $\Br^i$ and driven by a force $\BF^i$. 
We use bars to distinguish the functions associated with the constructed tensor
from those of the exact one. Since $\Bsigma^i$ is the exact stress for the case of a single particle, it is symmetric and divergenceless, $\Btau^i$ is symmetric and
traceless, and, thus, $\bar\Bsigma$ satisfies properties (a)--(c). In
addition, the integral of $\Bsigma^i$ over the perimeter of particle
$i$ is equal to $\BF^i$, whereas the integral of $\Bsigma^i$ over the
perimeter of another particle $j$ vanishes due to the divergence
theorem. Hence, $\bar\Bsigma$ satisfies requirement (d) as well. Therefore,
\begin{equation}
  \epsilon \leq \bar\epsilon = (2\mu)^{-1} \int d^2r\,\bar\Btau_{\alpha\beta}\bar\Btau_{\beta\alpha}
  =  \BF^i_\alpha \bar{\BB}^{ij}_{\alpha\beta}(\mathbbm{r})\BF^j_\beta,
\label{dissipation2}
\end{equation}
where $\bar{\BB}^{ij}_{\alpha\beta}$ is the approximate many-particle
mobility tensor (yet to be calculated), corresponding to the Ansatz 
(\ref{stress}).

The next step is to find the fields $\Bsigma^i$, $\Btau^i$, and $p^i$
induced by a single forced particle. They are readily obtained from
Saffman's treatment of the single-cylinder problem
\cite{Saffman}. The flow velocity $\Bu^i(\Br)$ at a point $\Br$ in the
fluid, resulting from a force $\BF^i$ applied to a single cylinder at
the origin, can be represented as
\begin{eqnarray}
&&\Bu_{\alpha}^i(\Br)=\BU_{\alpha \beta} (\Br) \BF_{\beta}^i \nonumber\\
&&\BU_{\alpha \beta}(\Br) =  \frac{1}{4 \pi \mu} \left[ \left(\ln \frac{2}{\kappa r}-\gamma-\frac{1}{2}+\frac{a^2}{2 r^2} \right)\Bdelta_{\alpha\beta}+\left(1-\frac{a^2}{r^2} \right) \frac{\Br_{\alpha} \Br_ {\beta}}{r^2} \right].
  \label{U}
  \end{eqnarray}
The resulting viscous stress and pressure, obtained from the
equations, $\Btau^i_{\alpha\beta} = \mu(\pd_\alpha\Bu^i_\beta +
\pd_\beta\Bu^i_\alpha)$ and $\pd_\beta\Bsigma^i_{\alpha\beta} =
-\pd_{\alpha}p^i + \pd_{\beta}\Btau^i_{\alpha \beta} = 0$, are
\begin{eqnarray}
  \Bsigma^i_{\alpha\beta}(\Br) &=& -p^i\Bdelta_{\alpha\beta} + \Btau^i_{\alpha\beta}  \nonumber \\
  \Btau^i_{\alpha\beta}(\Br) &=& \BT _{\alpha \beta \gamma} (\Br) \BF_{\gamma}^i,\ \ \ \ 
  p^i (\Br) = \frac{\Br_{\alpha} \BF_{\alpha}^i}{2 \pi \mu r},
\label{pressure}
\end{eqnarray}
where the tensor $\BT_{\alpha \beta \gamma}(\Br)=\mu \left( \partial_{\beta} \BU_{\alpha \gamma} + \partial_{\alpha} \BU_{\beta \gamma} \right)$. These results are substituted into Eq.~(\ref{stress}) to yield the
total stress $ \bar\Bsigma_{\alpha\beta}(\Br)$. Note that while Eq.~(\ref{pressure}) refers to the stress induced by an individual particle, the total stress in Eq.~(\ref{stress}) is a superposition of all such individual contributions.

The last step is to calculate the dissipation rate produced by the
constructed viscous stress and extract the approximate
mobility tensor [see Eq.~(\ref{dissipation2})]. The integral in
Eq.~(\ref{dissipation2}) contains diagonal terms, $I_1=\int d^2r\,
\Btau^i_{\alpha\beta}(\Br-\Br^i)\Btau^i_{\beta\alpha}(\Br-\Br^i)$, and
off-diagonal ones, $I_2=\int d^2r\,
\Btau^i_{\alpha\beta}(\Br-\Br^i)\Btau^{j\neq
  i}_{\beta\alpha}(\Br-\Br^j)$. The integration over the fluid area of the sheet, excluding the areas of the cylindrical objects, is technically very difficult. Therefore, repeating the arguments of Ref.~\citenum{RP}, we extend the definition of each individual stress $\Btau^i$ into the interior domain of the relevant particle $i$, such that $\Btau^i(|\Br-\Br^i| <a)=0$. The stresses emanating from particles $i$ and $j$ are not zero within the interior of particle $k~(k \neq i,j$). Thus, when we replace the correct integration domain with the whole area of the sheet, we introduce superfluous contributions to the dissipation from areas that do not contain a viscous fluid (for example, the contribution from the integral of $\Btau^1\Btau^2$ over the internal area of particle $3$). The sum of all these contributions is the integral of $\bar{\Btau}\bar{\Btau}$ over the non-fluid areas, which is strictly positive, and therefore further increases the dissipation rate beyond $\bar{\epsilon}$. Finally, we replace the integrals $I_1$ and $I_2$ by the following perimeter integrals:
\begin{eqnarray*}
&&I'_1=a \int_0^{2\pi} d\theta
  \, \Bu^i_\alpha(\Bw^i)\,\Bsigma^i_{\alpha\beta}(\Bw^i)\,
  \Bn^i_\beta = \BF^i_\alpha \bar{\BB}^{ii}_{\alpha\beta} \BF^i_\beta,\\
 &&I'_2=a \int_0^{2\pi} d\theta
  \, \Bu^i_\alpha(\Bw^i)\,\Bsigma^j_{\alpha\beta}(\Bw^j)\,
  \Bn^j_\beta = \BF^i_\alpha \bar{\BB}^{i\neq j}_{\alpha\beta} \BF^j_\beta.
\end{eqnarray*}
In these expressions $\Bw^{i,j}(\theta)=\Br^{i,j}+a\Bn^{i,j}$, where $\Bn^i(\theta)$ and $\Bn^j(\theta)$ are unit normals to the
perimeters of particles $i$ and $j$, using the same value of $\theta$
for the two along the integration. The divergence theorem and the facts that $\Bsigma$ is divergenceless and $\Btau$ is traceless allow us to transform the integrals $I'_{1,2}$
into $I_{1,2}$. The nonlocal terms, coupling the flow velocity at the perimeter of one particle
with the stress at the perimeter of another, are a nonintuitive
but direct consequence of the superposition Ansatz (\ref{stress}). Substituting
Eqs.~(\ref{U}) and (\ref{pressure}) into the integrals $I'_{1,2}$ and evaluating these
integrals, we extract the PDT,
\begin{eqnarray}
  \bar{\BB}^{i=j}_{\alpha\beta} &=& \frac{1}{4 \pi \mu}
  \left(\ln \frac{2}{\kappa a} -\gamma \right) \Bdelta_{\alpha\beta},
  \nonumber\\
  \bar{\BB}^{i\neq j}_{\alpha\beta} &=& \frac{1}{4 \pi \mu}
  \left [\left(\ln \frac{2}{\kappa r^{ij}} -\gamma -\frac{1}{2} +\frac{a^2}
    {(r^{ij})^2}\right)
    \Bdelta_{\alpha \beta} + \left(1-\frac{2 a^2}{(r^{ij})^2}\right)
    \frac{\Br^{ij}_{\alpha} \Br^{ij}_{\beta}}{(r^{ij})^2}\right],
\label{finaltensor}
\end{eqnarray}
where $r^{ij}=|\Br^{ij}|=|\Br^j-\Br^i|$. 

This equation is our central result. The diagonal terms simply
reproduce Saffman's self-mobility, Eq.~(\ref{Bs}). The off-diagonal
terms, as in the original RPY tensor for 3D suspensions, introduce
corrections of order $(a/r^{ij})^2$ to the 2D version of the KR
tensor, Eqs.~(\ref{Bpair}) and (\ref{Bpsd}).

In the appendices we present three elaborations on this
result. Appendix~A shows that Eq.~(\ref{finaltensor}) is equivalent to
a superposition of pair-mobilities, corrected for finite particle
size. Appendix~B extends the mobility tensor to cases where particles
may overlap ($r^{ij}\leq 2a$). In Appendix~C we suggest a useful
extension for large separations in fluid membranes
($r^{ij}>\kappa^{-1}$).

\subsection{Test case 1: Two cylinders}

\begin{figure}
 \centering
 \includegraphics[width=0.5\columnwidth]{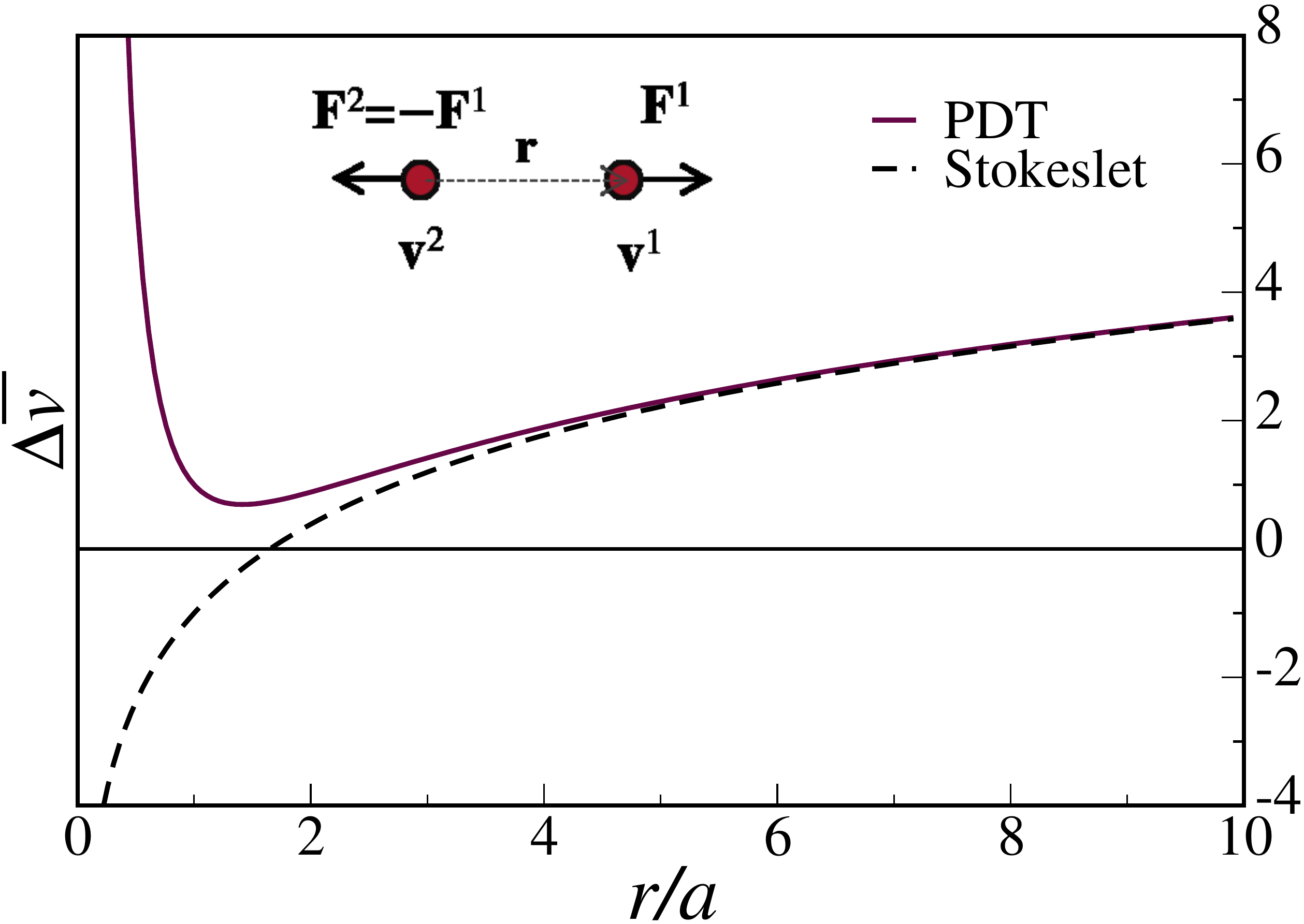}
\caption{Normalized velocity difference, $\Delta \bar{v}=4 \pi \mu
  \Delta v/F$ ($\Delta v=v^1-v^2$), of two cylindrical particles of
  radius $a$, separated by distance $r$ and pulled in opposite
  directions by forces of the same magnitude $F$, plotted versus the
  ratio $r/a$. The Stokeslet approximation (dashed black line) yields
  negative $\Delta v$ (particles move inward while driven outward)
  at the overlapping distances, $r \lesssim 1.65a$, while the PDT
  (maroon solid line) gives strictly positive results for all $r/a$
  values.}
\label{fig:2p}
\end{figure}

Let us consider the simplest configuration, an isolated pair of cylindrical particles (or membrane inclusions) separated by $\Br=\Br^1-\Br^2=r{\bf\hat{x}}$ and driven apart by the opposing forces $\BF^1=-\BF^2=F{\bf\hat{x}}$, as shown in Fig.~\ref{fig:2p}.

Following Eq.~(\ref{finaltensor}), the relative velocity of the particles is\footnote{As the distance $r$ approaches the cut-off $\kappa^{-1}$, the relative velocity saturates to the one between two noninteracting particles.}
\begin{equation}
\Delta v=v^1-v^2=\frac{F}{4 \pi \mu} \left( 2 \ln \frac{r}{a}-1+\frac{2 a^2}{r^2} \right).
\label{deltav}
\end{equation}
When the correction term $2a^2/r^2$ is neglected, the relative
velocity becomes negative for $r/a < \sqrt{e} \simeq 1.65$, meaning
that at these separations the particles get closer when pushed
apart. As shown in Fig.~\ref{fig:2p}, the correction term ensures
positive mobility at all separations. This test case produces a
maximum interaction effect, and, therefore, the mobility should remain
positive for any other choice of forces. In this simplest example of
two particles the problem arises at overlapping distances. In the next
example, however, the problem appears at larger, attainable
separations.

\subsection{Test case 2: Ring of normally driven cylinders}

\begin{figure}
 \centering
 \includegraphics[width=0.6\columnwidth]{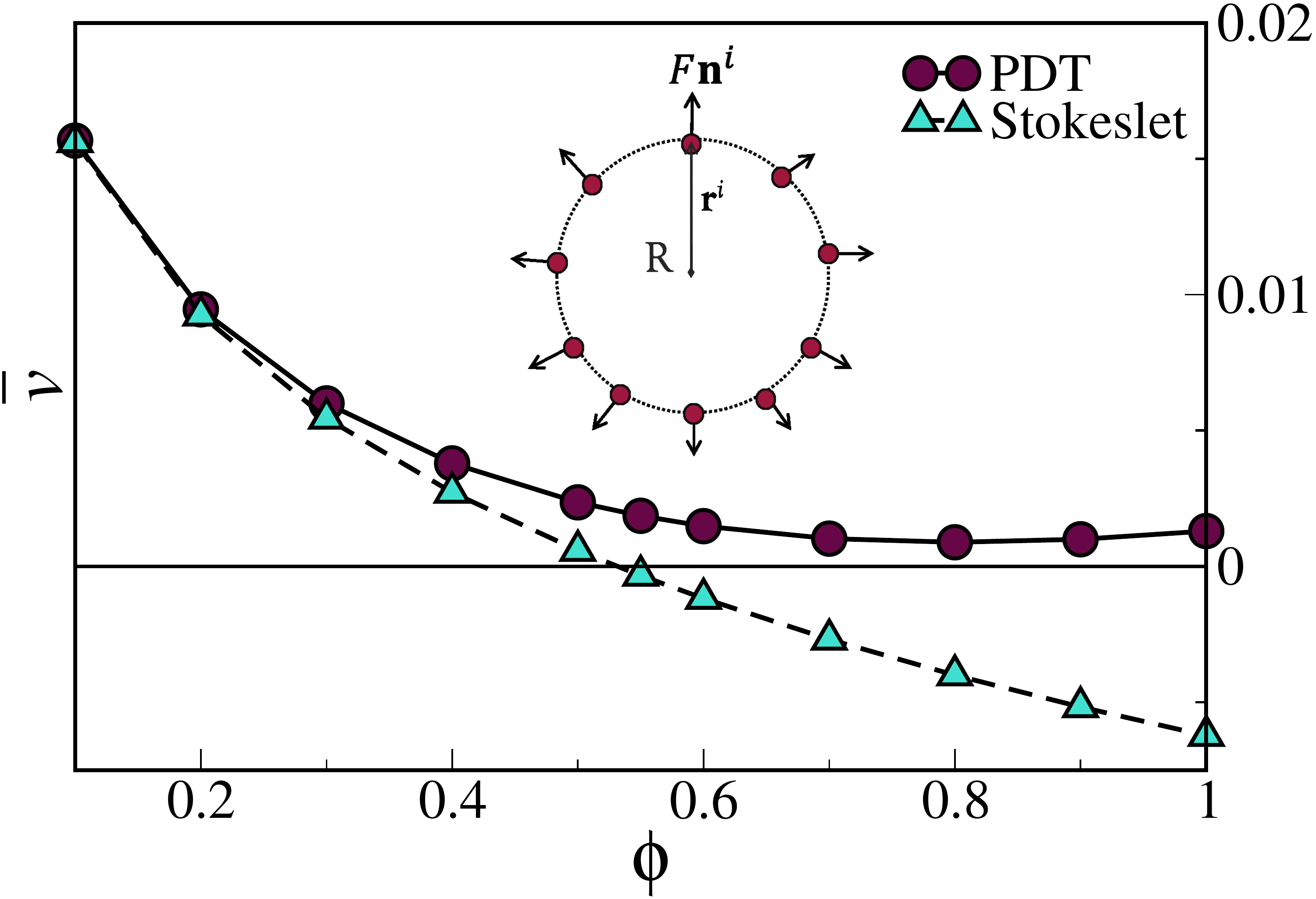}
\caption{Normalized velocities, $\bar{v}=4 \pi \mu v/F$, of $N=10$
  cylindrical particles of radius $a$, arranged in a symmetrical ring
  of radius $R$ and normally driven by a force of magnitude $F$,
  plotted versus particle density $\phi$ (line fraction). The
  velocities calculated using the Stokeslet approximation (blue
  triangles, dashed line) become negative above a certain density
  $\phi_c \simeq 0.55$. The velocities calculated using the
  positive-definite tensor (maroon circles, solid line) are positive
  at all densities. The normalized curves depend on $\kappa$ and $N$
  only, and we have used $\kappa=10^{-3}$. The value of $a/R$ is set
  at each point by the density, according to $a/R=\pi\phi/N$.}
\label{fig:ring}
\end{figure}

We now examine an initial configuration of $N=10$ cylindrical
particles evenly distributed along a circular ring of radius
$R=1$. The particles are driven in the normal outward direction by
identical forces $\BF=F\hat{\Bn}$ (see Fig.~\ref{fig:ring}) and
interact hydrodynamically. We take $\kappa=10^{-3}$ and $\mu=1$ and
check various line fractions $\phi$, such that $a=\phi \pi/N$. At
every time-step we calculate the velocity of each particle according
to Eq.~(\ref{finaltensor}) and advance them to their new positions at
the next step. The velocity hardly changes within the considered $50$
steps. We compare the results obtained with and without the correction
term (see Fig.~\ref{fig:ring}).

The figure shows that above a certain particle density, $\phi_c
\simeq 0.55$, the uncorrected tensor yields negative velocities ---
the ring shrinks under an outward forcing. For the same parameters,
the new tensor gives strictly positive velocities --- the ring always
expands, for all line fractions. As the density increases, the
calculated velocities, although positive, become inaccurate. In
particular, at $\phi=1$, the velocity is expected to vanish due to the
impossibility of inward flow, while the approximate tensor produces a
relatively small finite value.

\section{Discussion}
\label{sec_disc}

This work addresses the many-body hydrodynamic interactions among
driven or diffusing cylindrical objects within a viscous
sheet. Treating two test cases, a pair of particles and a ring of many
particles, we have demonstrated the elimination of non-physical
situations of negative mobility, using the derived PDT. Therefore, the
tensor can be safely used in numerical calculations involving
short-range hydrodynamic interactions, such as Stokesian Dynamics
schemes \cite{Stokesdyn} for viscous sheets.

Despite the ensured stability, we stress again that the formalism and
the resulting tensor are not exact, but limited to second order in the
ratio of particle size to separation. A treatment of many-body
dynamics which would be accurate to higher orders, should involve not
only the correction to the pair-mobility, but a departure from the
pair-wise superposition assumption as well.

One application of the formalism developed here is directly related to
the permeability of immobile protein assemblies in membranes to lipid
flow. This problem will be discussed in a forthcoming
publication. Another example would be an improved calculation of the
self-mobility or diffusivity of extended objects, such as rods and
polymer molecules, embedded in a viscous sheet or membrane
\cite{LevineMob,Noruzifar}. We recall that in such calculations, the
correction found by Rotne and Prager to the Kirkwood-Riseman tensor
vanishes upon spatial averaging, as noted by Yamakawa
\cite{Yamakawa}. The same holds for our correction to the Stokeslet
approximation in 2D. Therefore, the preaveraging approximation for
extended objects should not be used with the corrected tensor, as it
will nullify the effects of the correction term.

An important extension of this work would be to take into consideration viscoelastic (frequency-dependent) effects, either within the sheet \cite{Camley2}, or in its surrounding environment \cite{Komura}.

\begin{acknowledgments}
We thank Frank Brown and Naomi Oppenheimer for helpful comments. This
research has been supported by the Israel Science Foundation (Grant
No.~164/14).
\end{acknowledgments}

\appendix

\section*{Appendix A: Tensor derivation based on pair-mobility}

In this Appendix we recalculate the corrected pair-mobility of two
cylindrical particles following Ref.~\citenum{Naomi}. We show that the
many-particle tensor constructed from superposition of these
pair-mobilities coincides with the PDT derived in the main text.

The starting point is once again the flow field induced by a single
cylinder located at the origin and driven by the force $\BF^1$. Our
purpose is to calculate the velocity $\Bv^2(\Br)$ of another,
force-free, cylinder, embedded in this flow field at position
$\Br$. For a sphere in a 3D viscous fluid, this relation between the
particle velocity $\Bv^2(\Br)$ and the flow velocity $\Bu(\Br)$ is
given by Fax\'en's first law \cite{Faxen}. The analogous law for a
cylinder in a viscous sheet, in the limit $\kappa^{-1} \gg a$, was
derived in Ref.~\citenum{Naomi},
\begin{eqnarray}
\Bv^2(\Br)&=&\Bu(\Br)+\frac{a^2}{4} \nabla^2 \Bu.
\label{Lap}
\end{eqnarray}
Substituting Eq.~(\ref{U}) in Eq.~(\ref{Lap}), we obtain $\Bv_{\alpha}^2 (\Br)=\BB_{\alpha \beta}^{12}(\Br) \BF_{\beta}^1$ with the pair-mobility
\begin{equation}
\BB_{\alpha \beta}^{12}=\frac{1}{4 \pi \mu} \left[ \left(\ln \frac{2}{\kappa r} -\gamma -\frac{1}{2} +\frac{a^2}{r^2} \right) \Bdelta_{\alpha \beta} +
\left(1-\frac{2 a^2}{r^2} \right)\frac{\Br_{\alpha} \Br_{\beta}}{r^2}\right].
\label{Faxten}
\end{equation}
This result coincides with the pair-mobility block $\bar{\BB}^{i\neq
  j}_{\alpha\beta}$ of Eq.~(\ref{finaltensor}). Hence, the
many-particle mobility tensor, constructed from such pair-mobilities,
is identical to our PDT. Note, however, that this alternative
derivation does not prove positive-definiteness of the constructed
tensor.

\section*{Appendix B: Overlapping particles}

The mobility tensor derived in the main text, Eq.~(\ref{finaltensor}),
is guaranteed to be positive-definite provided that the disks do not
overlap, i.e., $r^{ij}>2a$ for all pairs $i,j$. (Its derivation relied
on the flow and stress fields around a single forced disk, which are
defined only outside the disk.) This restriction does not pose a
problem if short-range repulsion between particles is included, as is
done in most simulations. Still, extending the mobility tensor to
include overlapping separations may be useful, e.g., in simulations
where direct interactions are not of interest.  The ``inner'' tensor
(for $r^{ij}\leq 2a$) can be derived by recalculating the integrals in
Sec.~\ref{sec_PDT} while replacing the two circular boundaries of the
interacting disks by the boundary of two overlapping disks, as was
done for the 3D case in Refs.~\citenum{RP,Wajnryb2013}. Here we obtain
the inner tensor in a simpler way, by postulating its general form and
imposing the conditions that it should satisfy.

Based on the required integrations and the 3D
calculation,\cite{RP,Wajnryb2013} we anticipate an inner tensor
of the form,
\begin{eqnarray}
  \bar{\BB}^{i\neq j}_{\alpha\beta}(r^{ij}\leq 2a) = &&\frac{1}{4\pi\mu}
  \left[ \left( C_1 + C_2 \frac{r^{ij}}{2a} + C_3 \frac{r^{ij}}{2a}
    \ln \frac{r^{ij}}{2a} \right) \Bdelta_{\alpha\beta} \right.
    \nonumber\\
    &&\left. + \left( D_1 + D_2 \frac{r^{ij}}{2a} + D_3\frac{r^{ij}}{2a}
    \ln \frac{r^{ij}}{2a} \right)
    \frac{\Br^{ij}_{\alpha}\Br^{ij}_{\beta}}{(r^{ij})^2} \right].
\label{tensoroverlap}
\end{eqnarray}
To find the six coefficients we impose the following conditions
(satisfied also in the 3D case): (a) the inner tensor should be
divergenceless, similar to the outer one; (b) it should converge
continuously to the outer tensor at $r^{ij}=2a$; (c) at $r^{ij}=0$, as
the two disks overlap perfectly, they move together, i.e., the tensor
should converge to $\Bs\Bdelta_{\alpha\beta}$.  These conditions
yield,
\begin{eqnarray}
  C_1 &=& \ln \frac{2}{\kappa a} -\gamma, \nonumber\\
  C_2 &=& -\ln 2 - \frac{1}{4}, \nonumber\\
  C_3 &=& 2\ln 2 - \frac{3}{2}, \nonumber\\
  D_1 &=& 0, \nonumber\\
  D_2 &=& \frac{1}{2},\nonumber\\
  D_3 &=& - \ln 2 +\frac{3}{4}.
\label{coeffs}
\end{eqnarray}
Equations~(\ref{finaltensor}), (\ref{tensoroverlap}), and (\ref{coeffs}), 
together, give the mobility tensor for disks at all distances,
including the possibility of overlap.

\section*{Appendix C: Large separations in membranes}

In the main text, the tensor of Eq.~(\ref{finaltensor}) is limited to
particle separations smaller than the cutoff, $r^{ij}\ll\kappa^{-1}$
for all pairs $i,j$. In the particular case of fluid membranes, the
cutoff distance, and the hydrodynamic interaction beyond it, are well
characterized, arising from the coupling of the membrane to the
surrounding 3D fluid. It would be important for large-scale
simulations of membrane inclusions to have a mobility tensor which is
valid also for $r^{ij}\gtrsim\kappa^{-1}$.

A natural extension of Eq.~(\ref{finaltensor}) is the replacement of
the logarithmic Green's function, Eq.~(\ref{Bpsd}), by the full
Green's function for a membrane,\cite{Naomi,LevineDyn} while keeping
the short-range ($\sim a^2/r^2$) correction terms. This results in
\begin{eqnarray}
  \bar{\BB}^{i=j}_{\alpha\beta} = &&\frac{1}{4 \pi \mu}
  \left(\ln \frac{2}{\kappa a} -\gamma \right) \Bdelta_{\alpha\beta},
  \nonumber\\
  \bar{\BB}^{i\neq j}_{\alpha\beta}(r^{ij}) = &&\frac{1}{4 \mu}
  \left\{ \left[ H_0(\kappa r^{ij}) - \frac{H_1(\kappa r^{ij})}{\kappa r^{ij}}
     -\frac{1}{2}\left(Y_0(\kappa r^{ij}) - Y_2(\kappa r^{ij}) \right) +
     \frac{2 + a^2}{\pi(\kappa r^{ij})^2}     
    \right] \Bdelta_{\alpha \beta} \right. \nonumber\\
  &&- \left. \left[ H_0(\kappa r^{ij})-
    \frac{2H_1(\kappa r^{ij})}{\kappa r^{ij}} +
    Y_2(\kappa r^{ij}) + \frac{2(2+a^2)}{\pi(\kappa r^{ij})^2}
    \right]
    \frac{\Br^{ij}_{\alpha} \Br^{ij}_{\beta}}{(r^{ij})^2}\right\},
\label{tensormembrane}
\end{eqnarray}
where $Y_n$ and $H_n$ are, respectively, Bessel functions of the
second kind and Struve functions. This tensor covers all values of
$r^{ij}$, whether smaller or larger than $\kappa^{-1}$, as long as
$r^{ij},\kappa^{-1}\gg a$. For separations $r^{ij}\ll\kappa^{-1}$ it
coincides with the PDT of Eq.~(\ref{finaltensor}), while at larger
separations the positiveness issue is irrelevant. Thus, even though we
cannot provide a rigorous proof for the positive-definiteness of
Eq.~(\ref{tensormembrane}), this mobility tensor is very plausibly
positive-definite for all configurations of membrane inclusions.

\end{document}